\documentclass[%
 prl,longbibliography,twocolumn,notitlepage,
superscriptaddress,
 amsmath,amssymb,aps,
]{revtex4-2}

\usepackage{graphicx}% Include figure files
\usepackage{dcolumn}% Align table columns on decimal point
\usepackage{bm}% bold math
\usepackage{xcolor}
\usepackage{hyperref}% add hypertext capabilities
\usepackage{bbold}

\usepackage{soul}

\renewcommand {\phi}{{\varphi}}

\newcommand {\rmd}{{\rm d}}

\newcommand{\SigmaCon} {\Sigma^c}
\newcommand{\SigmaUn} {\Sigma^u}
\newcommand{\XiEx} {{\Xi^{ex}}}

\begin{document}

%\preprint{APS/123-QED}

\title{
 Nonequilibrium  entanglement between levitated masses under optimal control
}

\author{Alexander N. Poddubny}
\email{poddubny@weizmann.ac.il}
\affiliation{Department of  Physics of Complex Systems, Weizmann Institute of Science, Rehovot 7610001, Israel}

\author{Klemens Winkler}
\affiliation{University of Vienna, Faculty of Physics, Vienna Center for Quantum Science and Technology (VCQ), Boltzmanngasse 5, A-1090 Vienna, Austria}

\author{Benjamin A. Stickler}
 \affiliation{Institute for Complex Quantum Systems, Ulm University, Albert-Einstein-Allee 11, D-89069 Ulm, Germany}

 \author{Uroš Delić}

\affiliation{University of Vienna, Faculty of Physics, Vienna Center for Quantum Science and Technology (VCQ), Boltzmanngasse 5, A-1090 Vienna, Austria}

\author{Markus Aspelmeyer}
\affiliation{University of Vienna, Faculty of Physics, Vienna Center for Quantum Science and Technology (VCQ), Boltzmanngasse 5, A-1090 Vienna, Austria}
\affiliation{Institute for Quantum Optics and Quantum Information (IQOQI) Vienna, Austrian Academy of Sciences, Boltzmanngasse 3, 1090 Vienna, Austria
}

\author{Anton V. Zasedatelev}
\email{anton.zasedatelev@aalto.fi}
\affiliation{University of Vienna, Faculty of Physics, Vienna Center for Quantum Science and Technology (VCQ), Boltzmanngasse 5, A-1090 Vienna, Austria}
\affiliation{Department of Applied Physics, Aalto University School of Science, P.O. Box 15100, Aalto FI-00076, Finland}

\date{\today}

\begin{abstract}

We present a protocol that maximizes unconditional entanglement generation between two masses interacting directly through $1/r^{n}$ potential. The protocol combines optimal quantum control of continuously measured masses with their non-equilibrium dynamics, driven by a time-dependent interaction strength. Applied to a pair of optically trapped sub-micron particles coupled via electrostatic interaction, our protocol enables unconditional entanglement generation at the fundamental limit of the conditional state and with an order of magnitude smaller interaction between the masses compared to the existing steady-state approaches.

\end{abstract}

\maketitle
The progress in optomechanics~\cite{aspelmeyer2014cavity,barzanjeh2022optomechanics,GonzalezBallestero2021Levitodynamics} enables quantum state preparation of individual large-mass systems~\cite{teufel2011sideband,chan2011laser,Delic2020Cooling,Magrini2021RealTimeControl,Tebbenjohanns2021QuantumControl} and long-range cavity-mediated entanglement generation~\cite{OckeloenKorppi2018StabilizedEntanglement, Riedinger2018RemoteEntanglement}.
In addition to offering a valuable resource in quantum sensing and metrology~\cite{RevModPhys.90.025004,barzanjeh2022optomechanics}, entanglement between macroscopic systems holds conceptual significance for fundamental research. Being mediated via direct physical interaction through Coulomb, Casimir or the Newtonian force, it will facilitate the search for dark matter ~\cite{PhysRevLett.125.181102,kilian2024dark,Moore_2021}, the exploration of physics beyond the standard model ~\cite{RevModPhys.85.471}, and might eventually provide insights into the question of whether the gravitational interaction is fundamentally quantum or not ~\cite{dewitt2011role, Lami_2024}.

However, the cavity-free entanglement generation between two masses poses significant experimental challenges~\cite{Rudolph2022ForceGradientSensing,PhysRevLett.127.023601} and has yet to be achieved. Interactions with the environment rapidly destroy quantum correlations as the size and complexity of systems grows~\cite{RevModPhys.90.025004}, posing the necessity for practical methods to enhance quantum correlations against large decoherence rates. Strategies such as reservoir engineering which involves coupling a macroscopic systems to nonequilibrium baths~\cite{ludwig2010entanglement,estrada2015quantum} have been proposed to facilitate entanglement generation. Additionally, parametrically-driven interactions offer an innovative approach to achieve entanglement at higher temperatures, surpassing the need for deep cooling toward the ground state~\cite{Galve2010HighTemperatureEntanglement,szorkovszky2014mechanical,lin2020entangling}. By driving systems out of their thermal equilibrium, e.g. through a modulated interaction between subsystems, one can break the thermodynamic limit of the entanglement $k_{B}T\ll\hbar\omega$ (with $\hbar \omega$ being the typical energy of the system) ~\cite{mari2009gently,Galve2010HighTemperatureEntanglement}. While this methodology has primarily been developed for cavity optomechanical systems, recent theoretical work explores entanglement generation in free space through weak forces, suggesting protocols based on squeezing the motion of large masses~\cite{PhysRevLett.127.023601,PhysRevA.103.L061501}.

\begin{figure}
\includegraphics[width=0.45\textwidth]{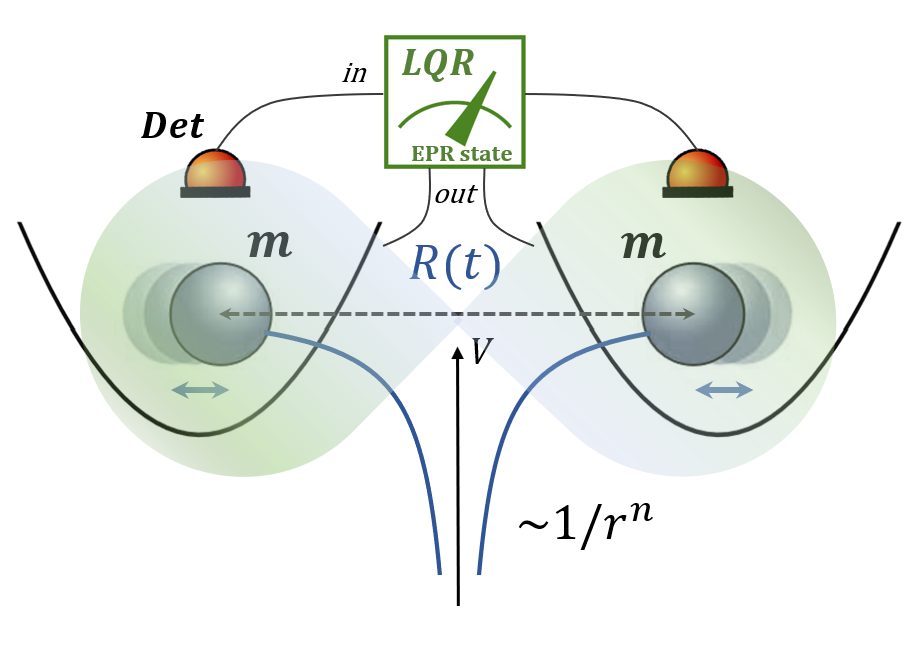}%Fig4TimeDep/FigAttractive4e.m
\caption{Schematic illustration  of the non-equilibrium entanglement generation between two levitating masses $m$, interacting via $\sim{1/r^n}$ potential in free space, subject to parametric modulation of the separation $\textit{R(t)}$. Both masses undergo continuous position measurement (\textit{Det}) and feedback control (\textit{LQR}) steering their motion to the entangled state.
}
%Fig4TimeDep/FigAttractive4d.m
    \label{fig:Fig1}
\end{figure}

In this work, we combine methods of nonequilibrium entanglement generation with the time-continuous quantum control~\cite{hofer2015entanglement} of two harmonically trapped masses interacting in free space. Figure~\ref{fig:Fig1} presents a schematic illustration of the system under consideration. We leverage continuous measurement and feedback with a periodic modulation of the interaction to drive the system into a non-equilibrium state that exhibits strong unconditional entanglement.%Our protocol features continuous interferometric measurement of their center-of-mass motion of two masses under phase-coherent modulation of the interaction strength between them.
We show how an optimal quantum measurement can be performed to generate entanglement in the conditional state~\cite{winkler2024stationary}. Following this strategy, we connect the observable entanglement to the properties of the measurement apparatus~\cite{hofer2015entanglement}. Then we use feedback control to stabilize the effective temperature of the motion, thus optimizing entanglement likelihood through a specifically chosen cost function. Ultimately, our protocol achieves unconditional (\textit{out-of-loop}) entanglement between the motion of directly interacting masses at the limit of their joint conditional state.
 %%%%%%%%%%%%%%%%%%%%%%%%%%%%%%%%%%%%%%%%%%%%
\begin{figure}
\includegraphics[width=0.5\textwidth]{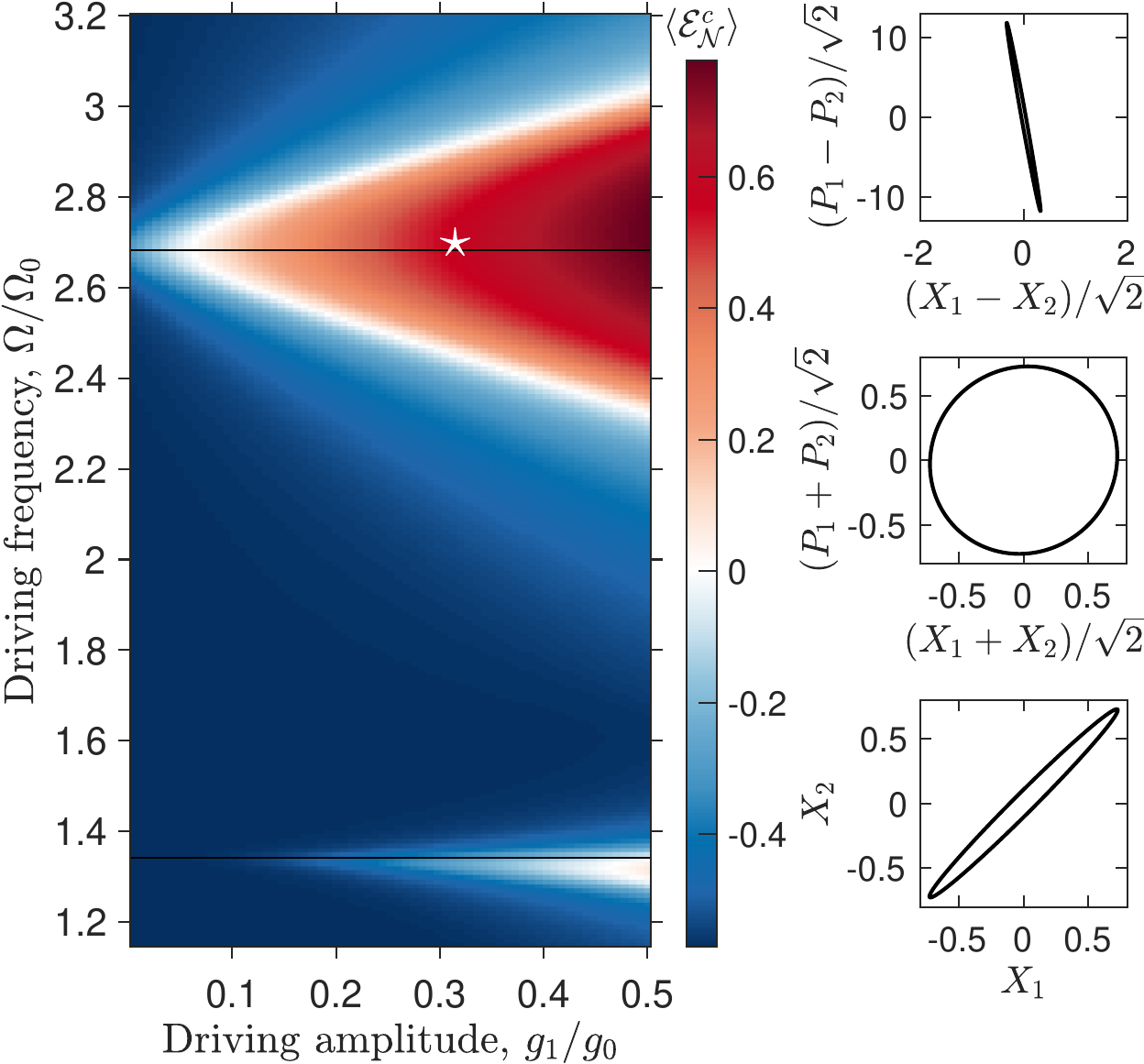}%FigResonances/Fig1_revised2025April7
\caption{Conditional period-averaged logarithmic negativity as the function of driving amplitude $g_1$ and modulation frequency $\Omega$. Panels on the right illustrate the standard deviation of the bivariate Wigner distribution, given by the conditional covariance matrix $\SigmaCon$, depending on coordinates and momenta of the symmetric (\textit{common}) and antisymmetric (\textit{differential}) modes $X_1\pm X_2$ and $P_1\pm P_2$ for a characteristic value of $\{g_1/g_0,\Omega/\Omega_0\}=\{0.3,2.7\}$, labeled by a star on the color plot. Here we use 
the following parameters of the system: $g_0/\Omega_0=0.2$, $\Gamma_{\rm ba}/\Omega_0 = 5\%$, $\Gamma_{\rm th}/\Gamma_{\rm ba}=5\%$ and $q= 0.1/\Omega_0$, fully consistent with the parameters in Ref.~\cite{winkler2024stationary}.
%\textcolor{red}{The calculation has been carried out for $P=1.98\times 10^{-10}~$mBar, $\varphi_{\rm EPR}=\pi$, independent control, $R=52.4~$nm, $g_0/\Omega_0=0.2$, $q=5.41\times 10^{-7}s$, \blue{$\eta=0.5$}. This results in $\Gamma_{\rm ba}/\Omega_0 = 5\%$, $\Gamma_{\rm th}/\Gamma_{\rm ba}=5\%$ and $q= 0.1/\Omega_0$, consistent with the parameters chosen in \cite{winkler2024stationary}.}
%The calculation has been carried out for $P=1.71\times 10^{-10}~$mBar, $\varphi_{\rm EPR}=\pi$, independent control, $R=52.4~$nm, $g_0/\Omega_0=0.2$,$q=1.08\times 10^{-6}$, \blue{$\eta=0.5$}. 
}
    \label{fig:Fig2}
\end{figure}
%%%%%%%%%%%%%%%%%%%%%%%%%%%%%%%%%%%%%%%%%%%%

We consider two identical particles with mass $m$, each confined in one spatial dimension by a harmonic potential with eigenfrequency  $\Omega_0$ as shown in Fig.~\ref{fig:Fig1}. The particles are modeled to interact via a central potential of the form $k/r^n$, e.g. via gravitation, electrostatic or Casimir interaction, where we introduce $r= \sqrt{x_{\rm zpf}^2\left(X_1-X_2\right)^2+R^2}$ with particle positions $X_{k}$. Here, $R$ denotes the interparticle distance along a spatial direction orthogonal to the motion $X_{1,2}$ with $x_{\rm zpf} = \sqrt{\hbar/m\Omega_0}$. In the limit of small displacements %, i.e. $R/x_{\rm zpf}\gg \langle X_i\rangle$, 
%the interaction Hamiltonian can be expanded into  a Taylor series. Considering the terms up to the quadratic order in position 
and considering terms up to quadratic order in position yields the interaction Hamiltonian $H_{\rm int} = \hbar g (X_1-X_2)^2$ with a coupling rate $g=k/(2 n R^{2+n}m\Omega_0)$. For the following discussion, we assume that the coupling rate can be temporally modulated with the frequency $\Omega$ e.g. by varying the distance $R$, such that 
$g(t)=g_0+2g_1\cos\Omega t$. Additionally, the particles are controlled individually by a feedback force $u_k$. The total Hamiltonian is then given by
\begin{equation}\label{eq:Hamiltonian}
\begin{gathered}
     H= \sum_{k=1,2} \frac{\hbar\Omega_0}{2}\left(X_k^2+ P_k^2\right) -\sum_{k=1,2} \hbar X_ku_k(t) \\
     +\hbar g(t) (X_1-X_2)^2.
\end{gathered}
\end{equation}
% In the rotating reference frame, the parametric  modulation of the coupling $g(t)$ leads to the term 
% $\propto (a_1^\dag a_2^\dag-a_1a_2)$
% \cite{Wineland1990,Tian_2008}, where $a_{1,2}$ are the annihilation operators for the oscillator motion. This term describes two-mode squeezing and entanglement generation.

The interaction term acts selectively on the differential mode leading to quadrature squeezing $\sim \hbar g(t) \left( \hat{a}_-^2 + \hat{a}_-^{\dagger 2} \right)$, where $\hat{a}_- = (\hat{a}_1 - \hat{a}_2)/\sqrt{2}$ is the differential-mode annihilation operator, defined via the bare ladder operators $\hat{a}_1$ and $\hat{a}_2$ of the two oscillators. When the time-dependent interaction strength hits the resonant frequency \( \Omega = 2\Omega_- \), the squeezing is further enhanced $\sim g_1$, leading to the buildup of strong non-classical correlations in the out-of-phase motion of the particles and resulting in entanglement generation (see Supplementary Materials for detailed analysis).

%We consider relevant decoherence channels, including thermal decoherence from particle interaction with residual gas  and back-action decoherence from photon recoil with rates $\Gamma_{\rm th}$ and $\Gamma_{\rm ba}$ respectively. 
Relevant decoherence channels include thermal decoherence from residual gas and photon recoil with rates $\Gamma_{\rm th}=\gamma n_{\rm th}$ \cite{Magrini2021RealTimeControl}  and $\Gamma_{\rm ba}$ respectively. Here, the mean thermal occupation number $n_{\rm th}$ quantifies the thermal energy of the environments each particle is in contact with, and is given in the high-temperature limit by $n_{\rm th}\approx k_{\rm B}T/{\hbar \Omega_0}$. These settings are typical for various continuously measured interferometric arrangements, such as optical trapping of sub-micron particles in ultra-high vacuum, which offers extreme isolation, scalability, and potential for landscape engineering~\cite{GonzalezBallestero2021Levitodynamics}.

Continuous position measurement of each particle, e.g. via homodyne detection, conditions the system on the measured photocurrents, leading to the conditional state represented by the density matrix $\rho_c$. With Gaussian initial states and the linear dynamics following from  Eq.~\ref{eq:Hamiltonian}, the conditional state remains Gaussian, fully described by its first and second moments.

%With each particle undergoing continuous position measurement e.g. via independent homodyne detection, the system is being conditioned on the measured photocurrents, giving rise to the corresponding conditional state represented by the density matrix $\rho_c$. Assuming Gaussian initial states and with the linear dynamics generated by the Hamiltonian from \eqref{eq:Hamiltonian}, the state will remain Gaussian and is fully characterized by the conditional first and second moments.

%Introducing the 4-vector of dimensionless coordinates \(X_{1,2}\) and momenta \(P_{1,2}\) quadratures of the two particles, 
Using \(X_{1,2}\) and dimensionless momenta \(P_{1,2}\) we define $\mathbf{X}=(X_1,P_1,X_2,P_2)^T$, the conditional mean values $\mathbf{X}_c := \text{Tr}[\mathbf{X}\rho_c] = \langle \mathbf{X} \rangle_c$ and the covariance matrix $\SigmaCon_{kl} = \langle X_k X_l \rangle_c - \langle X_k \rangle_c \langle X_l \rangle_c$.

Following the notations from \cite{winkler2024stationary}, with the time-modulated drift matrix $\mathbf{A}(t)$ 
\begin{equation}\label{eq:drift-mat}
\mathbf A(t)=\begin{pmatrix}
0&\Omega_0&0&0\\
-\Omega_0-2g(t)&-\gamma&2g(t)&0\\
0&0&0&\Omega_0\\
2g(t)&0&-\Omega_0-2g(t)&-\gamma
\end{pmatrix}.
\end{equation}
the evolution of the systems conditional mean values and covariance matrix is described by
\begin{subequations}
\begin{align}
d\mathbf{X}_c(t) &= \mathbf{A}(t) \mathbf{X}_c(t)dt + \mathbf{B}\mathbf{u}(t)dt\notag\\
&+ \SigmaCon(t)\mathbf{C}^T \mathbf{W}^{-1}d\mathbf{w}(t)\label{eq:cond_mean}\:,\\
d\SigmaCon &= \mathbf{A}(t)\SigmaCon(t)dt + \SigmaCon(t)\mathbf{A}(t)^T dt + \mathbf{V}dt\notag\\
&-\SigmaCon(t)\mathbf{C}^T\mathbf{W}^{-1}\mathbf{C}\SigmaCon(t)dt\label{eq:cond_cov}
\end{align}
\end{subequations}
with control  matrix $\mathbf{B}$, input force vector $\mathbf{u} = (u_1, u_2)^T$, measurement matrix $\mathbf{C}$, system and measurement noise correlation matrices $\mathbf{V}$ and $\mathbf{W}$ and stochastic vector $d\mathbf{w}(t)=\left(dW_1(t), dW_2(t)\right)^T$. %with $\mathbb{E}[dW_k(t)dW_l(t)]=\delta_{kl}dt/2$ \cite{winkler2024stationary}. 
All presented matrices are displayed in their full form in the \textit{end matter} of this work.% \ref{supp:matrices}. 

Making the interferometric measurement of the position quadrature unavoidably perturbs the system, causing it to explore a larger phase space. Subject to excess noise, the averaged conditional state turns into the unconditional state of the system \(\rho = \mathbb{E}[\rho_c]\)~\cite{hofer2015entanglement}. The covariance matrix of the unconditional state is given by the sum of the conditional state and the excess noise \(\SigmaUn = \SigmaCon + \XiEx\). Since $\XiEx,\SigmaCon$ are positive definite, the conditional state occupies a smaller region in phase space compared to the unconditional state. As the conditional state is fully determined by the measurement setup and independent of feedback, it imposes a fundamental limit on the entanglement of the unconditional state for a given finite interaction rate\cite{winkler2024stationary}.

To minimize excess noise, we apply the theoretical framework of classical control theory with the Kalman-Bucy filter as  optimal observer together with a linear quadratic regulator (LQR) as optimal controller. The Kalman-Bucy filter provides a recursive state estimation by balancing information from system dynamics and measurement, incorporating the underlying state-space evolution, including process and measurement noise. The LQR, formulated as an independent optimization problem, determines the optimal feedback strategies for linear systems by   minimizing the quadratic cost function
$\mathcal{J} = \int \mathbb{E}[ \mathbf{X}_c^T \mathbf{P} \mathbf{X}_c +  \mathbf{u}^T q \mathbf{u} ] dt$, 
 where  \(\mathbf{P}\) is the cost matrix, and  \(q\) represents the control effort \cite{stengel1994optimal,xiong2008introduction}. The LQR relies on the state estimation provided by the Kalman-Bucy filter but operates independently in determining the optimal feedback forces for the system. 
The implementation of these feedback forces depends on the specific feedback setup and can be realized through various mechanisms, such as electric forces via electrodes \cite{Magrini2021RealTimeControl}, optical forces e.g. by shifts in the central trap position or parametric feedback by using  intensity modulation of the trapping laser \cite{Vijayan_2023, Reisenbauer_2024}, or  magnetic forces \cite{Tian_2024}. Regardless of the chosen approach, it is beneficial for both particles to be independently addressable by feedback forces, as discussed in \cite{winkler2024stationary}.

From now on, we will use the Einstein-Podolsky-Rosen (EPR) cost function, which minimizes EPR-type variances in the joint phase space of the two particles, as %presented in the supplementary material %Z\ref{supp:matrices} and 
detailed in Ref. \cite{winkler2024stationary}. 
The optimal feedback $\mathbf{u}(t) = -\mathbf{K}(t)\mathbf{X}_c(t)$ is determined by the feedback gain $\mathbf{K}(t) = \mathbf{B}^T\Omega(t)/q$, solving the backwards Riccati equation
\begin{equation}
d\Omega(t)
= -[\mathbf{A}^T(t) \Omega(t)+\Omega(t) \mathbf{A}(t)+ \mathbf{P}]dt +\Omega(t)\frac{\mathbf{B} \mathbf{B}^T}{q}\Omega(t)dt.
\end{equation}
The excess noise matrix $\XiEx$ %in the unconditional covariance 
depends on both, the measurement setup and the feedback from the controller, its time evolution is governed by ~\cite{hofer2015entanglement}
\begin{align}\label{eq:Excess_noise}
d\XiEx(t)=&
     \left(\mathbf{A}-\mathbf{B}\mathbf{K}(t)\right)\XiEx(t)dt \\\notag
     +& \XiEx(t) \left(\mathbf{A}-\mathbf{B}\mathbf{K}(t)\right)^Tdt \\\notag
     +& \left(\SigmaCon(t) \mathbf{C}^T+\mathbf{M}\right)\mathbf{W}^{-1}\left(\SigmaCon(t) \mathbf{C}^T+\mathbf{M}\right)^Tdt\:. 
\end{align}

Defining $\tilde{\nu}$ as the minimal symplectic eigenvalue of the either the matrix $\SigmaCon$ or $\SigmaUn$, we employ the function $ \mathcal{E}_\mathcal{N}=  -\text{ln} (2\tilde{\nu})$ to quantify the conditional and unconditional entanglement of the state respectively. States are detected as entangled if $\mathcal{E}_\mathcal{N}>0$, connecting to the the logarithmic negativity, a measure for entanglement, via $\rm max\left(0,\mathcal{E}_\mathcal{N}\right)$ ~\cite{Adesso_2007, Werner_2002}.

%This function is connected to the logarithmic negativity, being an entanglement measure ~\cite{Adesso_2007, Werner_2002}, by , states are detected as entangled if  $\mathcal{E}_\mathcal{N}>0$. 

%We employ the logarithmic negativity $ \mathcal{E}_\mathcal{N}=  -\text{ln} (2\tilde{\nu})$ to characterize the entanglement measure of the state~\cite{Adesso_2007, Werner_2002}, where $\tilde{\nu}$ is the minimal symplectic eigenvalue of the either the matrix $\SigmaCon$ or $\SigmaUn$ for conditional and unconditional  entanglement. States are detected as entangled if  $\mathcal{E}_\mathcal{N}>0$. Here, we allow the logarithmic negativity to yield negative values for separable states to get a better understanding of the entanglement dynamics.

\begin{figure}
\includegraphics[width=\linewidth]{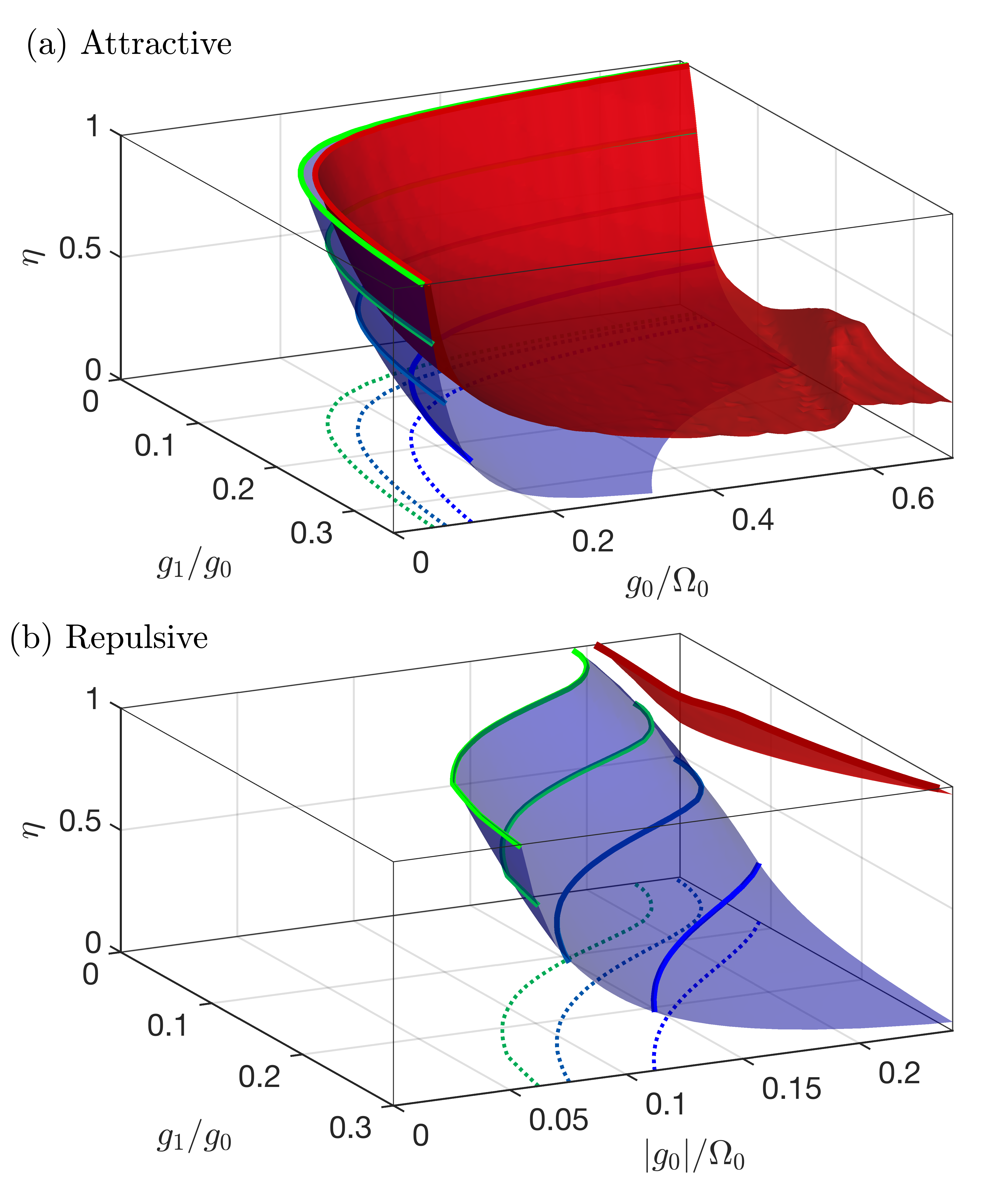}%Fig3v2.m
\caption{
The boundary for the conditional and unconditional time-averaged entanglement is shown by the blue
and dark red surfaces, respectively.
The case of attractive interaction $g_0>0$ (a) and repulsive interaction $g_0<0$ (b).
Calculation has been performed for the EPR phase $\theta=\pi$ (a) and $\theta=0$ (b)
. %other parameters the same values of other parameters as in Fig.~\ref{fig:Fig2}
%$\varphi=0$. 
The contour curves (solid) show the boundaries for unconditional entanglement calculated at $\eta=0.25,~0.5,~0.75$ and~$1$, with their projections illustrated by the dotted curves. 
}
%Fig3Repulsive/Fig3Rep.m
    \label{fig:Fig3}
\end{figure}

%%%%%%%%%%%%%%%%%%%%%%%%%%%%%%%%%%%%%%%%%%%%%%%
% \begin{figure}
% \includegraphics[width=0.48\textwidth]{Fig4C}
% \caption{
% Time-dependent logarithmic negativity $E(t)$ for
% (a) attractive case with $g_0/\Omega_0=0.2$ and 
% (b) repulsive case with $g_0/\Omega_0=-0.2$.
% For each case we show the data for three value of $g_1/g_0$ indicated on graph. Solid curves correspond to the conditional negativity, dotted curves correspond to the unconditional one. Other parameters are the same as in Fig.~\ref{fig:Fig3}.
% Panels (c--f) illustrate the Wigner functions of conditional (solid curves) and unconditional (dotted curves) states, calculate at the time corresponding to maximal unconditional negativity.
% %FigAttractive4D.m
% }
%     \label{fig:Fig4C}
% \end{figure}
%%%%%%%%%%%%%%%%%%%%%%%%%%%%%%%%%%%%%%%%%%%%%%%

We solve the periodic time-dependent Riccati equations using the Schur decomposition method~\cite{Hench1994NumericalIntegration}.   In this approach, Eq.~\eqref{eq:cond_cov} is described by the Hamiltonian matrix
$\mathbb H=\left(\begin{smallmatrix}
    \mathbf A^T &- \mathbf C\mathbf W^{-1}\mathbf C\\ -\mathbf V&-\mathbf A
\end{smallmatrix}\right)
$.  Next, we evaluate the Schur decomposition of the evolution operator over the modulation period $\mathbb S=\exp(\int_0^{2\pi/\Omega} \mathbb H(t)\rmd t)$, $\mathbb S=\mathbb U \Lambda \mathbb U^T$, where $\Lambda$ is an upper-diagonal matrix and $\mathbb U=\left(\begin{smallmatrix} 
\mathbf U_{11}&\mathbf U_{12}\\\mathbf U_{21}&\mathbf U_{22}\end{smallmatrix}\right)$ is the unitary one.
The covariance matrix is then given by
$\hat\Sigma^c(t)=\mathbf U_{12}(t) [\mathbf U_{11}(t)]^{-1}$. 
The brute-force evaluation of the $\mathbb S$ matrix 
leads to severe numerical errors for low values of control effort $q$ and low modulation frequencies, since $\mathbb H(t)$ has eigenvalues with positive real part. 
This is in contrast to the Floquet engineering of closed quantum systems~\cite{Bukov_2015,Eckardt2017,Rudner2020,rudner2020floquet}. Although similar in form to $\mathbb{S}$, Floquet Hamiltonians predominantly govern the unitary evolution of a wavefunction, thereby avoiding such issues.
%There, one also typically introduces a Floquet Hamiltonian similar to $\mathbb S$, but  since the wavefunction evolution is unitary, no such errors arise.
One workaround is provided by the so-called periodic QR decomposition~\cite{Bojanczyk1992,Kressner_2006,Varga_2008}. In our experience, however, evaluating the $\mathbb{U}(t)$ dependence iteratively proved to be more numerically efficient~\cite{ding2015periodic}.
%%%%%%%%%%%%%%%%%%%%%%%%%%%%%%%%%%%%%%%%%%%%%%%
\begin{figure}
\includegraphics[width=0.48\textwidth]{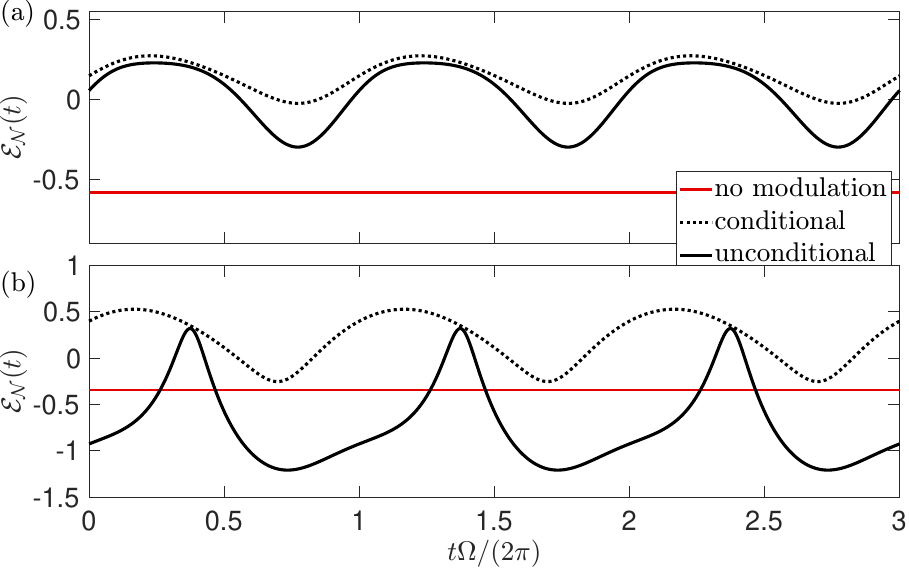}%Fig4TimeDep/FigAttractive4e.m
\caption{Time-dependent logarithmic negativity $\mathcal E_\mathcal N(t)$ at detection efficiency $\eta= 0.5$ for 
(a) attractive case with $g_0/\Omega_0=0.2$ and 
(b) repulsive case with $g_0/\Omega_0=-0.2$.
 Dotted black  curves correspond to the conditional negativity, solid black curves correspond to the unconditional one. Horizontal solid lines  show the  values of
$\mathcal E_\mathcal N(t)$ without the modulation, $g_1=0$.
Calculated for $|g_1/g_0|=0.15$, $\Omega/\Omega_0=2.74$, EPR phase $\theta=\pi$ (a) and 
%Fig4b corresponds to |g_0|=0.20, Fig 4b corresponds to |g_0|=0.22
$\Omega/\Omega_0=0.7$, $\theta=0$ (b).
}
%Fig4TimeDep/FigAttractive4d.m
    \label{fig:Fig4}
\end{figure}

%%%%%%%%%%%%%%%%%%%%%%%%%%%%%%%%%%%%%%%%%%%%%%%

According to the stationary theory under the attractive interaction $g_0>0$~\cite{Rudolph2022ForceGradientSensing}, the motion of particles remains separable unless the interaction strength exceeds the mechanical frequency $g_0\gtrsim\Omega_0$. Tailoring feedback control  to maximize entanglement generation can help to lower the separability threshold. However, this approach still necessitates $g_0\sim\Omega_0$~\cite{winkler2024stationary}, which unfortunately stretches far beyond current experimental reach ~\cite{Rieser2022TunableInteraction}. Therefore, realistic protocols should demonstrate a separability threshold at much lower interactions. We examine our protocol setting $g_0=0.2\Omega_0$, with the other parameters being the same as those considered in the stationary theory~\cite{winkler2024stationary}.

Figure~\ref{fig:Fig2} shows the conditional entanglement measure, logarithmic negativity $\langle\mathcal E_N^c \rangle$, calculated as a function of the amplitude $g_1$ and frequency $\Omega$ of the modulation. Red color regions correspond to the entangled conditional state,  where $\langle\mathcal E_N^c \rangle>0$. This happens at the parametric resonances, with the  main resonant frequency  being $\Omega=2\Omega^*$, $2\Omega^*=2\sqrt{\Omega_0^2+4\Omega_0g_0}\approx 2.74 \Omega_0$ for the chosen parameters. When the squeezing in the joint phase space is strong enough quantum correlations survive thermal fluctuations destroying it otherwise.  Increasing the modulation depth $g_1$ expands the entangled region. Furthermore, for sufficiently large  $g_{1}$, an additional parametric resonance emerges at the half of the drive frequency $\Omega=\Omega^*$. This result is well-known from the stability analysis of quantum systems under parametric drives~\cite{roque2013role}. 
The two-mode squeezing can also  be achieved through parametric driving of the particle frequencies using the correct phase relation $\phi$ between the frequency drives $\Omega_m(t)=\Omega_0 +\Omega_1 \cos(\Omega_{p} t+\phi)$. In the \textit{end matter} of this work, we also explore the potential of this parametric driving, discussing the results and the prospects of combining this approach with our current strategy of temporal coupling modulation to enhance the overall entanglement generation.
However, a modulation of the interaction strength naturally ensures the perfect phases for optimal squeezing, which is evidenced in Fig.~\ref{fig:Fig2}(i-iii).  The panels (i--iii)  illustrate the standard deviation of the corresponding Wigner function, characterized by the covariance matrix $\SigmaCon$ of the conditional state, calculated for the symmetric and anti-symmetric state with position and momentum quadrature $X_1\pm X_2$ and $P_1\pm P_2$ respectively as well as the Wigner function of $X_1$ versus $X_2$. 

Having established conditional entanglement generation via parametric modulation, we now explore a broader parameter space. The key parameters are the stationary interaction strength $g_0/\Omega_0$, the amplitude $g_1/g_0$ and frequency $\Omega/\Omega_0$ of the time-dependent interaction, and the detection efficiency $\eta$ of the time-continuous position measurement.  To present the calculated 
$\langle\mathcal E_N(g_0, g_1, \eta,
\Omega) \rangle$ within a single plot, for every combination of $\{g_0, g_1, \eta\}$, we select the maximum logarithmic negativity at the frequency $\Omega$, corresponding to the strongest parametric resonance in the system. Figure~\ref{fig:Fig3} shows the separability boundaries $\langle\mathcal E_N^c \rangle=0$ obtained in the case of attractive (a) and repulsive (b) interaction. Here, we represent the performance of both conditional (blue surface) and unconditional states (red surface). We also plot contours on the surface of the conditional state for  $\eta$ equal to 0.25, 0.5, 0.75, and 1, along with projections on the $(g_0,g_1)$ plane.
In the attractive regime with relatively low modulation depth $g_1=0.2g_0$ and realistic detection efficiency of $\eta=0.5$, our approach predicts entanglement generation at $g_0=0.1\Omega_0$. This is more than an order of magnitude lower in comparison to existing stationary protocols~\cite{Rudolph2022ForceGradientSensing,winkler2024stationary}. Stationary entanglement in the repulsive case requires $|g_0|>0.2\Omega_0$~\cite{winkler2024stationary}, but the system turns unstable when $|g_0|\geq0.25\Omega_0$, which makes it extremely hard to implement. The parametric drive of the interaction modifies limits of entanglement generation  fundamentally set by the conditional state, making it possible under more relaxed experimental conditions, Fig.~\ref{fig:Fig3}b.

The generation of time-averaged unconditional (\textit{out-of-loop}) entanglement demands much higher detection efficiency with the given parameters and is likely to remain unattainable in the experiments. Nonetheless, optimal quantum control that minimizes the EPR cost function enables robust generation of unconditional entanglement in the stroboscopic regime, establishing quantum correlations beyond the separability criteria at a specific time. This protocol relies on the continuous time-resolved phase space tomography, fully consistent with the introduced measurement-based feedback control.

Figure~\ref{fig:Fig4} demonstrates the time dependence of the logarithmic negativity of unconditional states (black solid curves) for the attractive (a) and repulsive (b) regimes. For comparison, we also include the results without the modulation (red dotted lines) and the negativity of conditional states with modulation (black dotted curves). We observe a significant increase in the logarithmic negativity of the unconditional state, surpassing the separability boundary $\mathcal E_N^{u}=0$, and approaching the fundamental limit of nonequilibrium entanglement expected for the conditional state. This is a remarkable result that allows for \textit{out-of-loop} entanglement generation between large masses under realistic experimental conditions. Importantly, the peaking period of the unconditional entanglement is persistent with the period of parametric resonance $1/\Omega$. Similar phenomena have been observed in amplitude-modulated optomechanical systems, where the entanglement  between the mirror motion and the cavity field oscillates with the period of the parametric drive~\cite{mari2009gently}.

\textit{Summary and outlook}~---~We have developed a protocol for unconditional entanglement generation between two masses in free space interacting via $1/r^{n}$ potential. The protocol involves continuous monitoring and optimal state estimation via Kalman filtering as well as LQR feedback to enhance the entanglement between two oscillators under parametrically driven interaction. We developed feedback control strategy specifically tailored to maximize entanglement generation. It is achieved by the cost function of the linear quadratic regulator which minimizes EPR-variances of the particles' motion, as detailed in Ref.\cite{winkler2024stationary}. The feedback loop generates a non-equilibrium unconditional entanglement, significantly reducing the complexity for future experiments.

% The parametric drive of interaction strength, resonant with the normal modes, optimally conditions the joint phase space for two-mode squeezing. This results in time-oscillating entanglement that peaks at instances of maximal squeezing. Our protocol significantly enhances entanglement generation over stationary methods, crossing the separability threshold under more relaxed experimental conditions and reaching the fundamental limit of the entanglement generation in the unconditional state. 
% We showcase its performance studying the Coulomb interaction between two sub-micron optically trapped particles. 
Our protocol achieves unconditional entanglement generation at the fundamental limit of the conditional state, ultimately making experimental implementation of the \textit{out-of-loop} entanglement fairly accessible in laboratories. In particular, operating in an attractive regime with a gentle modulation depth of 20\% and a realistic detection efficiency of 50\%, we achieve entanglement  at the interaction strength \(g_0=0.1\Omega_0\)~--~more than an order of magnitude improvement over existing stationary protocols~\cite{Rudolph2022ForceGradientSensing,winkler2024stationary}. 
These findings emphasize the potential of combining non-equilibrium dynamics with quantum control techniques to advance entanglement generation between macroscopic systems and setting the stage for future research with gravitational or Casimir interactions.
%We have developed a protocol for unconditional entanglement generation between two masses in free space interacting via 1/r potential. Our protocol takes advantage of nonequilibrium dynamics driven by parametrically modulated interparticle interaction and real-time optimal quantum control over the joint phase space of the motion. We showcase the performance of our protocol using the Coulomb interaction between two sub-micron optically trapped particles as the testbed. The protocol achieves unconditional entanglement generation between two large masses at the fundamental limit of the conditional state ultimately making experimental implementation of the \textit{out-of-loop} entanglement fairly accessible in laboratories. 
%Our results underscore the potential of combining non-equilibrium dynamics with quantum control techniques to explore new frontiers in quantum physics, setting foundation for future studies in the entanglement generation between macrsoscopic systems mediated by gravitational or Casimir interactions. 

\begin{acknowledgments}

The authors thank Andreas Deutschmann-Olek, Henning Rudolph, Ayub Khodaee, Nancy Gupta, Nikolai Kiesel, Gerard Higgins and Corentin Gut for insightful  discussions. 
K.W. and M.A. received funding from the European Research Council (ERC) under the European Union’s Horizon 2020 research and innovation program (grant agreement No 951234), and from the Research Network Quantum Aspects of Spacetime (TURIS). 
BAS acknowledges funding by the DFG–510794108 as well as by the Carl-Zeiss-Foundation through the project QPhoton.
U.D. acknowledges funding from the Austrian Science Fund (FWF, Project DOI 10.55776/I5111).
A.V.Z. acknowledges support from the European Union’s Horizon 2020 research and innovation programme under the Marie Sklodowska-Curie grant LOREN (grant agreement ID: 101030987). 
\end{acknowledgments}

%\nocite{apsrev41Control}
%\bibliographystyle{apsrev4-1}
%\bibliographystyle{apsrev4-2}
\bibliography{main}% Produces the bibliography via BibTeX.

\clearpage

\onecolumngrid
\begin{center}
\textbf{\large End Matter}\label{sec:end}
\end{center}

%%%%%%%%%% Merge with supplemental materials %%%%%%%%%%
%%%%%%%%%% Prefix a "S" to all equations, figures, tables and reset the counter %%%%%%%%%%

\setcounter{equation}{0}
\setcounter{figure}{0}
\setcounter{table}{0}
\setcounter{page}{1}
\makeatletter
\renewcommand{\theequation}{S\arabic{equation}}
\renewcommand{\thefigure}{S\arabic{figure}}
\renewcommand{\bibnumfmt}[1]{[S#1]}
\renewcommand{\citenumfont}[1]{S#1}

\section{Matrices for time evolution of conditional and unconditional state}\label{supp:matrices}

We present the full form of the matrices introduced for the time-evolution of the conditional means  $\mathbf{X}_c$ and covariances $\SigmaCon$ covered in %(3a) and (3b) 
 \ref{eq:cond_mean} and \ref{eq:cond_cov} 
 of the main text. For the considered system we define  the control matrix $\mathbf{B}$
\begin{equation}
        \begin{gathered}
        \mathbf{B} = \begin{pmatrix}
            0 & 0 & 0 & 1\\
            0 & 1 & 0 & 0
        \end{pmatrix}^T \label{eq:control_matrix}
    \end{gathered}
\end{equation}
which quantifies how the input forces $\mathbf{u}=(u_1, u_2)^T$ couple to the conditional coordinates and momenta. 
Assuming each particle motion is driven by its own force, the system is fully \textit{controllable} in terms of control theory \cite{Wiseman_2009, Bechhoefer_2021}.

Based on the efficiency $\eta$ for read-out of the particles motion with $0<\eta\leq1$, we introduce the  measurement rate $\Gamma_m=\eta\Gamma_{\rm ba}$ and the measurement matrix  $\mathbf{C}$ 
\begin{equation}
        \begin{gathered}
           \mathbf{C} =
           \sqrt{\Gamma_{\rm m}}\begin{pmatrix}
               1& 0 & 0 & 0\\
               0 & 0 & 1& 0 
           \end{pmatrix}
    \end{gathered}
\end{equation}
where we  assume individual continuous position measurement of both particles, given by the vector of measurement currents $\mathbf{I}dt=\left(I_1(t), I_2(t)\right)^Tdt = \mathbf{C}\mathbf{X}_cdt +d\mathbf{w}(t)$  \cite{Hofer_2015, Winkler_2024}, that makes the whole system to be \textit{observable} in terms of measurement theory \cite{Wiseman_2009, Bechhoefer_2021}.

As evident from %(3a)
Eq.~\ref{eq:cond_mean}
, the conditional means are driven by the stochastic vector $d\mathbf{w}(t) = \left(dW_1(t) , dW_2(t)\right)^T$ where the independent Wiener increments satisfy  $\mathbb{E}[dW_k(t)dW_l(t)]=\delta_{kl}dt/2$. In contrast, the time-evolution of the conditional matrix $\SigmaCon$ in Eq.~%(3b) 
\ref{eq:cond_cov} 
is entirely deterministic.  
We define the correlation matrices  $\mathbf{V}$ and $\mathbf{W}$ for the system and measurement noise, respectively, as follows: 
% \begin{subequations}
%     \begin{gather}
%     % \mathbf{V}= \left(\Gamma_{\rm ba}+\Gamma_{\rm th}\right)\begin{pmatrix}
%     %     0 & 0 & 0 & 0\\
%     %     0 & 1 & 0 & 0\\
%     %     0 & 0 & 0 & 0\\
%     %     0 & 0 & 0 & 1\\
%     % \end{pmatrix}\\
%          \mathbf{V} = \left(\Gamma_{\rm ba}+\Gamma_{\rm th}\right)\text{\rm diag}\left(0,1, 0,1\right)\\
%        \mathbf{W} = \frac{1}{2}\mathbb{1}_{2\times2}. 
%     \end{gather}
% \end{subequations}
\begin{equation}
    \mathbf{V} = \left(\Gamma_{\rm ba}+\Gamma_{\rm th}\right)\text{\rm diag}\left(0,1, 0,1\right)\quad
       \mathbf{W} = \frac{1}{2}\mathbb{1}_{2\times2}.
\end{equation}

We note that any non-singular transformation on the measurement currents $\mathbf{I}dt\mapsto\mathbf{OI}dt$, represented by the $2\times 2$ matrix $\mathbf{O}$, will not change the conditional means and covariances as it is evident from Eq.~\ref{eq:cond_mean} \& Eq.~\ref{eq:cond_cov} considering that

\begin{equation}
    \mathbf{W}\mapsto \mathbf{OWO}^T, \quad \mathbf{C}\mapsto\mathbf{OC}, \quad d\mathbf{w}(t)\mapsto\mathbf{O}d\mathbf{w}(t)\:.
\end{equation}
As a result, the conditional state, as well as the associated entanglement properties,  remain unaffected by a particular linear combination of position quadratures $X_1$ and $X_2$ detected.

In the main text, the linear quadratic regulator (LQR) is introduced to provide an optimal control force minimizing the cost function $\mathcal{J}=\int \mathbb{E}\left[\mathbf{X}_c^T \mathbf{P}\mathbf{X_c} +\mathbf{u}^Tq\mathbf{u}\right]dt$. Here, the control effort $q$ and control matrix $\mathbf{P}$ are chosen to resemble an Einstein-Podolsky-Rosen (EPR) variance with EPR-angle $\theta$. The EPR-cost function maximizes likelihood of entanglement generation, for a detailed discussion see \cite{Winkler_2024}: %, commonly employed as an entanglement witness for Gaussian states \cite{Duan_2000,Simon_2000}
\begin{equation}
\mathbf{P}= \Omega_0\begin{pmatrix}
    \mathbb{1}_{2\times2} & \mathbf{R}(\theta) \\
    \mathbf{R}(\theta) & \mathbb{1}_{2\times2}    
\end{pmatrix}\:, \quad \mathbf{R}(\theta) = \begin{pmatrix}
     \cos \theta& \sin \theta\\
    \sin \theta& -\cos \theta\\
\end{pmatrix}.
    % \mathbf{P}=\Omega_0 \begin{pmatrix}
    %     1 & 0 & \cos \theta& \sin \phi\\
    %     0 & 1 & \sin \theta& -\cos \theta\\
    %     \cos \theta& \sin \theta& 1 & 0\\
    %     \sin \theta& -\cos \theta& 0 & 1\\
    % \end{pmatrix}
\end{equation}

\section{Parametric Driving of particle frequencies}\label{supp:para_drive}

This section explores the impact of parametric driving of particle frequencies on generating conditional entanglement in the system. In the main text, we show that dynamically tuning the interactions between particles can substantially enhance the entanglement of motion.  However, the introduction of parametric frequency modulation adds a new layer of control \cite{Ballestero_2021} that is known to enhance correlations of motion leading to quadrature squeezing \cite{Mari_2009} .

Our analysis is applied to systems of optically trapped particles, where the trapping potential - and consequently the mechanical frequencies - can be modulated through the power $P$ of the trapping laser  \cite{Rashid_2016,Setter_2019}, as was recently employed in ultra-high vacuum \cite{Rossi_2024}. 
For two identical particles with radius $r$, density $\rho$ and electrical susceptibility $\chi$  placed in the focus of two independent Gaussian beams  with wavevector $k$ and Rayleigh length $x_R$, the  trapping frequency $\Omega_m$ of the center of mass motion of each particle is given by $\Omega_m(t) = \sqrt{\chi k P(t)/\pi c \rho x_R^3}$ \cite{Isart_2019}.
% \begin{equation}
%     \Omega_m(t) = \sqrt{\frac{\chi k P(t)}{\pi c \rho x_R^3}}. 
% \end{equation}

Photon scattering on the particles into free space introduces decoherence with back-action rate $\Gamma_{\rm ba}\propto\Omega_m(t)$ \cite{Jain_2016}. %\textbf{I'd suggest one of the earlier papers instead, e.g.Vijay Jain, Phys. Rev. Lett. 116, 243601 2016 }\cite{Henning_2021}.
Consequently,  modulating the laser power will lead to time-dependent variations in the back-action decoherence and the measurement rate   $\Gamma_{\rm m}$ while keeping the readout efficiency $\eta = \Gamma_{\rm m}/\Gamma_{\rm ba}$ constant over time. 
When parametric driving of mechanical frequencies is introduced through laser power modulation,  the control matrix $\mathbf{B}$ remains unaffected. However, the measurement matrix $\mathbf{C}$ along with the system and measurement noise correlation matrices   $\mathbf{V}$ and $\mathbf{W}$ become time-dependent. Additionally, the bare frequency $\Omega_0$ in the drift matrix $\mathbf{A}$ defined in (2) %\ref{eq:drift-mat} 
is replaced by the time-dependent frequency $\Omega_m(t)$. Despite these changes, the evolution of the conditional state is still fully described by (3a) and (3b).% \ref{eq:cond_mean} and \ref{eq:cond_cov}.

\begin{figure}
    \centering
    \includegraphics[width=0.45\linewidth]{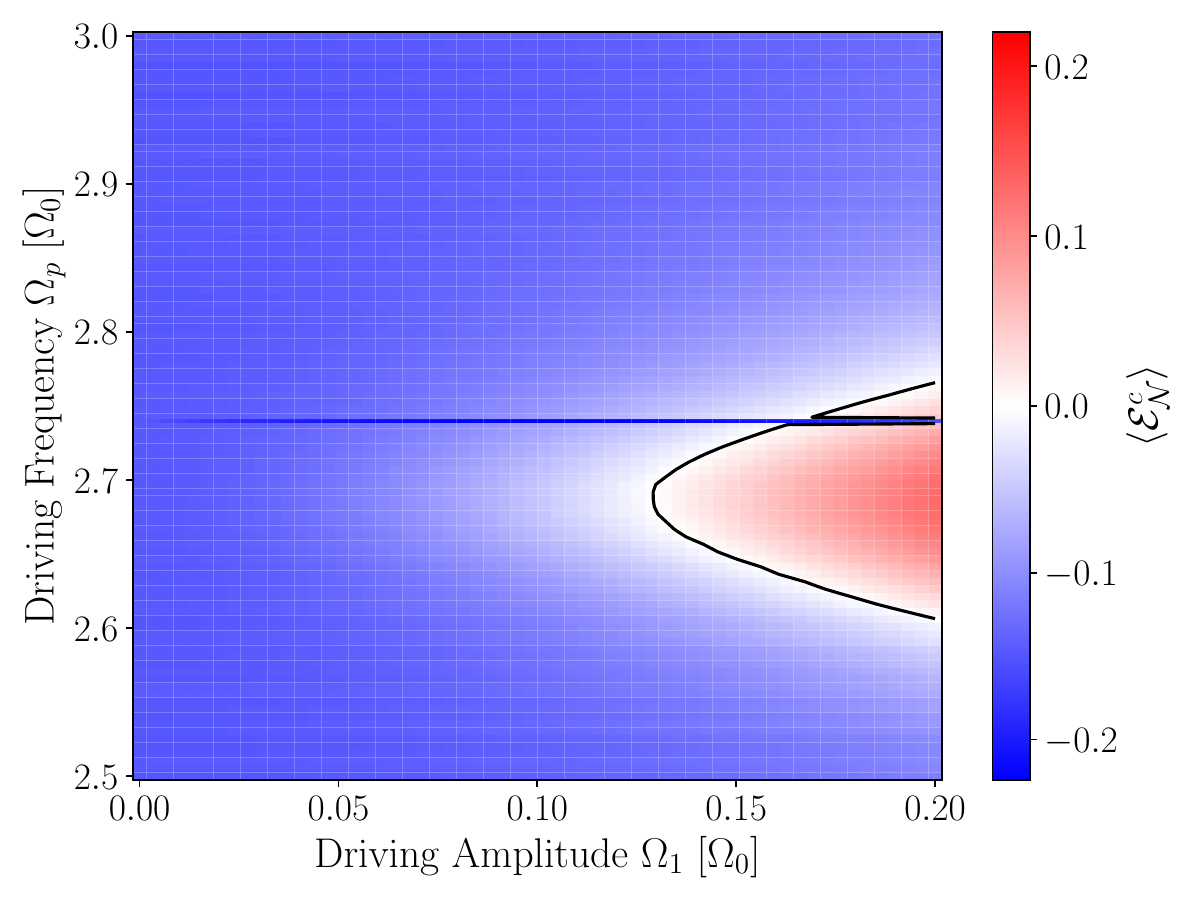}
    \caption{Time averaged logarithmic negativity $\mathcal{E}_\mathcal{N}^c(t)$ of the conditional state as a function of the in-phase ($\phi = 0$) parametric driving amplitude $\Omega_1$ and frequency $\Omega_p$, the solid black line represents the separability boundary between entangled and separable states. Here, we use the following parameters: $g_0/\Omega_0 = 0.2$, $g_1/\Omega_0 = 0.01$ and $\Omega/\Omega_0=2.74$, with  $\Gamma_{\rm ba }/\Omega_0 =5\%$ and $\Gamma_{\rm th}/\Gamma_{\rm ba} = 5\%$ matching those in  Fig.1 of the main text for $t=0$. }
    \label{fig:end-para}
\end{figure}

With control over the laser power, we modulate the mechanical frequency of both particles using a driving amplitude $\Omega_1$, driving frequency $\Omega_p$ and phase $\phi$  resulting in  $ \Omega_m(t) = \Omega_0 +\Omega_1\cos (\Omega_p t +\phi)$. To simulate the time evolution of the conditional state, we numerically integrate Eq.~(3b) %\ref{eq:cond_cov} 
and then extract time-averaged values of logarithmic negativity $\mathcal{E}_\mathcal{N}^c$, which characterizes the entanglement in the conditional state. The results are presented in Fig. \ref{fig:end-para}, showing $\mathcal{E}_\mathcal{N}^c$ plotted against the driving amplitude $\Omega_1$ and driving frequency $\Omega_p$ at phase $\phi=0$. The temporal modulation of the interaction strength is set to $g_0/\Omega_0=0.2$, $g_1/\Omega_0=0.01$ and $\Omega/\Omega_0 = 2.74$, with the decoherence rates matching those for Fig.1 in the main text at $t=0$: $\Gamma_{\rm ba }/\Omega_0 =5\%$ and $\Gamma_{\rm th}/\Gamma_{\rm ba} = 5\%$. 

Importantly, we do not expect entanglement in the conditional state under this configuration without modulating the frequencies, as shown in Fig.~\ref{fig:Fig2}. The results in Fig.~\ref{fig:end-para} clearly indicate that additional modulation applied to the mechanical frequency with $15\%$ of its bare value $\Omega_0$ in the vicinity of the parametric resonance $\sim2.7\Omega$ enables conditional entanglement generation between the particles. Therefore, controlling the mechanical frequencies introduces additional degrees of freedom to evolve the motion of both particles into an entangled state. We believe that a carefully selected combination of parametric driving of the mechanical frequency, along with modulation of the interaction strength under optimal quantum control of the joint phase space, paves the way for entanglement generation in realistic experiments.

\end{document}